\documentclass{jssc}

\def\textsubscript#1%
{$_{\text{#1}}$}

\def\cdd{\mbox{\boldmath$\cdot$}~}

\usepackage{graphicx}
\usepackage{mathtools}
\usepackage{amsfonts}
\usepackage{amssymb}
\usepackage{graphics}
\usepackage{color}
\usepackage{pgfplots}
\usepackage{hyperref}

\usepackage{url}
\usepackage{graphicx}
\usepackage{multirow}
\usepackage{multicol}
\usepackage{amssymb}
\usepackage{mathrsfs}
\usepackage{enumerate}
\usepackage{enumitem}
\newsavebox{\measurebox}

\usepackage{rotating}

\makeatletter
\def\@oddfoot{\hfill}
\newcount\shumeicount
\def\setshumei#1#2#3{%
  \shumeicount=\count0
  \def\@oddhead{%
    \raise-5pt\hbox to0pt{\vrule width\hsize height 0pt depth 0.4pt\hss}\relax
    \ifnum \shumeicount=\count0
      \raise-7pt\hbox to0pt{\vrule width\hsize height 0pt depth 0.4pt\hss}\relax
      #1
    \else
      \ifodd\count0
        #2
      \else
        #3
       \fi
     \fi
  }%
}
\makeatother
\makeatletter
\def\@oddfoot{\hfill}
\newcount\shujiaocount
\def\setshujiao{%
  \shujiaocount=\count0
  \def\@oddfoot{%
      \ifodd\count0
      \else
      \fi
  }%
}
\makeatother
\def\title#1#2#3#4{{
  \vspace*{0.3cm}
  \begin{flushleft} \Large\bf #1\end{flushleft}
  \vspace*{-0.2cm}
      \begin{flushleft}
      \bf #2
      \end{flushleft}
      \footnotetext{\hspace{-6mm} #3\\ #4}}}

\def\dshm#1#2#3#4
{\setshumei{J Syst Sci Complex (#1) #2:
{\thepage--\pageref{LastPage}}\hfill}
            {\hfill {\small #3}\hfill\hbox to0pt{\hss\thepage}}
            {\hbox to0pt{\thepage\hss}\hfill {\small #4}\hfill
            }
            \setshujiao}
\def\drd#1#2
{{\vskip 1cm\small \begin{flushleft}
 #1 \\
 #2\\
\copyright The Editorial Office of JSSC \&  Springer-Verlag GmbH
Germany 2018
\end{flushleft}}}

\def\epsilon{\varepsilon}

\begin{document}

\title{Affinity Classification Problem by Stochastic Cellular Automata}
{Kamalika \uppercase{Bhattacharjee}  \cdd Subrata \uppercase{Paul}
 \cdd Sukanta \uppercase{Das} }
{Kamalika \uppercase{Bhattacharjee}  \\
Department of Computer Science and Engineering, National Institute of Technology, Tiruchirappalli, Tamilnadu - 620015. Email: kamalika.it@gmail.com \\   
Subrata \uppercase{Paul}
 \cdd Sukanta \uppercase{Das}  \\
Department of Information Technology, Indian Institute of Engineering Science and Technology, Shibpur, P.O. - Botanic Garden, Howrah, West Bengal-711103.  Email:	subratapaul.sp.sp@gmail.com; sukanta@it.iiests.ac.in
   } 
{}

\drd{DOI: }{Received: x x 20xx}{ / Revised: x x 20xx}


\dshm{20XX}{XX}{Affinity Classification Problem by Stochastic Cellular Automata}{\uppercase{Bhattacharjee Kamalika} $\cdd$ \uppercase{Paul Subrata} $\cdd$ \uppercase{Das Sukanta}
}

\Abstract{This work introduces a new problem, named as, \emph{affinity classification problem} which is a generalization of the density classification problem. To solve this problem, we introduce \emph{temporally stochastic} cellular automata where two rules are stochastically applied in each step on all cells of the automata. Our model is defined on $2$-dimensional grid having \emph{affection} capability. We show that this model can be used in several applications like modeling self-healing systems.}      



\section{Introduction}
\noindent The density classification problem is a well-known problem in cellular automata (CAs). Given an initial configuration, this problem asks to find a binary cellular automaton (CA) that converges to all-$0$ (resp. all-$1$) configuration, a fixed point, if number of $0$’s (resp. $1$s) in the initial configuration in higher than the number of $1$s (resp. $0$s). That is, the CA reaches all-$1$ configuration if it has an \emph{affinity} towards $1$ in its initial configuration with respect to the density of $1$ in it and reaches all-$0$ otherwise. However, sometimes, the requirement of many applications is that, this density itself is to be treated as a variable -- still a binary CA is required that can converge to the all-$1$ (resp. all-$0$) configuration. In this paper, we introduce this problem as a generalization of the density classification problem. Formally the problem can be stated as:\\\\
\textbf{Problem Statement:} \emph{Given an initial configuration, find a binary cellular automaton that converges to all-$1$ configuration if density of $1$s is more than $\rho$. Otherwise, it converges to all-$0$ configuration.}\\\\
Here, $\rho$ is calculated as the density of $1$s in the initial configuration and all-$0$ and all-$1$ are the only fixed points of the CA. We name this problem as \emph{Affinity Classification Problem} as the CA has an affection towards the all-$1$ configuration. When $\rho = 0.5$, the problem is reduced to the classical density classification problem.

In literature, several attempts have been taken to solve the density classification problem. However, in \cite{PhysRevLett.74.5148}, it is proved that it is impossible to solve this problem with $100\%$ accuracy using classical CAs. Because of this, research efforts have been shifted towards finding the non-classical CAs which can solve the problem {\em almost} perfectly.
In \cite{Fuk05},  it is shown that the density classification task is solvable by running in sequence the trivial combination of elementary rules $184$ and $232$. This solution is extended for two-dimension using a stochastic component into each of these two rules in \cite{fuks2015solving}. In \cite{fates00608485}, a stochastic CA is used to solve the problem with an arbitrary precision. In this solution, the cells of 1-dimensional CA stochastically choose a rule in each step from a set of rules to evolve. These non-classical CAs can be named as \emph{spatially} stochastic CAs. Target has also been taken to tackle this problem with non-uniform CA where the cells can use different rules to evolve. A non-uniform CA that does the best density classification task is identified in \cite{NazmaTh}. However, neither (spatial) stochastic CA nor non-uniform CA can perfectly solve the density classification problem. Whereas, the non-classical CA of Ref.\cite{Fuk05} which may be called as \emph{temporally} non-uniform CA, can do it perfectly. 

As the affinity classification problem is an extension of the density classification problem, it is most likely to be unsolvable using classical CAs. We may need non-classical CA with temporal non-uniformity and stochastic component for this. Hence, to solve this problem, in this work, we introduce \emph{temporally stochastic} CAs. We define our problem over two dimensional binary CAs and use two different CA rules uniformly over the grid. The default rule is deterministic, whereas, another rule is stochastic whose application time is dependent on some probability. Section~\ref{model} describes the proposed model. The simulation and convergence to the solution for different density is shown in Section~\ref{simulation}. It is shown that our model is not blind as it \emph{intelligently} decides and converges to its \emph{point of attraction}.
Finally, we show that this model has several applications including as model for \emph{self-healing systems} (Section~\ref{application}).

\section{The Model}\label{model}

\noindent The proposed cellular automaton is defined over two-dimensional square grid which uses periodic boundary condition. The CA is binary and considers Moore neighborhood dependency; that is, a cell takes any of the two states $0$ or $1$ and depends on itself and it's eight nearest neighbors. At a time stamp $t$, a cell can be updated using one of the two rules $f$ and $g$. Here, $f$ is deterministic and the default rule for the grid, whereas, $g$ is stochastic and is applied with some probability. As the CA is defined over Moore neighborhood, both $f$ and $g$ are having the same domain and range:
$$f:\{0,1\}^9 \rightarrow \{0,1\} \text{            and            } g:\{0,1\}^9 \rightarrow \{0,1\}$$

Let us now first discuss about the default rule $f$. This rule is spatially deterministic -- at any time, it is applied over all cells uniformly. At each time step $t+1$, this rule updates the state of cell ${(i,j)}$ depending on the present states of its neighboring cells:

\begin{scriptsize}
	$${(i-1,j)}, {(i-1,j-1)}, {(i,j-1)},  {(i+1,j-1)}, {(i+1,j)}, {(i+1,j+1)}, {(i,j+1)}, {(i-1,j+1)}$$
\end{scriptsize}

Let $s^t_{i,j}$ be the present state of cell ${(i,j)}$ and $\mathscr{C}^d_{(i,j)}$ represents for the cell $(i,j)$, $s^t_{i,j}=d$ where $d \in \{0, 1\}$.
Then $f$ works in the following way:
\begin{align*}
s^{t+1}_{i,j} & = f(s^t_{i-1,j}, s^t_{i-1,j-1}, s^t_{i,j-1} ,  s^t_{i,j} , s^t_{i+1,j-1}, s^t_{i+1,j}, s^t_{i+1,j+1} ,  s^t_{i,j+1} , s^t_{i-1,j+1})\\
& = \begin{cases}
0 & \text{ if } s^t_{i,j}=1 \text{ and } \sum\limits_{\substack{i-1 \le l \le i+1,\\ j-1 \le m \le j+1}}\mathscr{C}^0_{(l,m)}> K  \\
1 & \text{ if } s^t_{i,j}=0 \text{ and } \sum\limits_{\substack{i-1 \le l \le i+1,\\ j-1 \le m \le j+1}}\mathscr{C}^1_{(l,m)}=8-K\\
s^t_{i,j} & \text{ otherwise }
\end{cases}
\end{align*}
where $K$ is a constant and $0\le K \le 8$. 
That means, if a cell is $1$ and it has more than $K$ neighbors with state 0, it becomes 0 in next step; whereas, a cell of state 0 with $(8-K)$ or more neighbors with state 1 becomes 1 in the next step. This number of neighbors required for state transition ($K$) is the \emph{first parameter} of the model.

The most significant characteristics of our model comes from the second rule $g$. As already mentioned, $g$ is a stochastic rule, that is, it is applied to each cell with some probabilities. Moreover, at which time step this rule is to be applied that is also stochastically decided. Hence, we call the CA as a temporally stochastic CA. However, when selected, this rule is also applied uniformly over all cells. Following is the definition of this rule:
\begin{align*}
s^{t+1}_{i,j} & = g(s^t_{i-1,j}, s^t_{i-1,j-1}, s^t_{i,j-1} ,  s^t_{i,j} , s^t_{i+1,j-1}, s^t_{i+1,j}, s^t_{i+1,j+1} ,  s^t_{i,j+1} , s^t_{i-1,j+1})\\
& = \begin{cases}
0 \text{ with probability } \phi(x)& \text{ if } s^t_{i,j}=1 \text{ and } \sum\limits_{\substack{i-1 \le l \le i+1,\\ j-1 \le m \le j+1}}\mathscr{C}^0_{(l,m)}=x \\
1 \text{ with probability } \psi(x) & \text{ if } s^t_{i,j}=0 \text{ and } \sum\limits_{\substack{i-1 \le l \le i+1,\\ j-1 \le m \le j+1}}\mathscr{C}^1_{(l,m)}=x\\
s^t_{i,j} & \text{ otherwise }
\end{cases}
\end{align*}
Here, $\phi(x)$, $\psi(x): \{0,1,\cdots, K\}\rightarrow [0,1]$ are two probability distribution functions.
We denote this $x$ as the number of \textit{supporting neighbors} or simply \emph{support}.

This rule implies, if a cell is at state $1$ and it has $x$ number of neighbors with state $0$, it updates its value to $0$ with some probability $\phi(x)$. Similarly, if a cell is at state $0$ and it has $x$ number of neighbors with state $1$, it updates its value to $1$ with some probability $\psi(x)$. We name $\phi(x)$ as the \emph{affection probability} and $\psi(x)$ as the \emph{repulsion probability} function. These two probability distribution functions are the \emph{second} and \emph{third parameters} of our model.

However this stochastic rule $g$ does not act in each step. When it is to be applied is decided by another probability $p$, which we name as the \emph{upgrade probability}. This $p$ is the \emph{fourth} and final \emph{parameter} of our model. Hence, the parameters required by the model are --
\begin{itemize}
	\item $K$ = number of neighbors required to change from one state to another
	\item $\phi(x)$= affection probability function
	\item $\psi(x)$ = repulsion probability function
	\item $p$ = upgrade probability
\end{itemize}

Observe that, in our model the role of $g$ is to give the cells an extra chance to change their status. During evolution of the CA by $f$ if some cells are left out which are \emph{eager} to update their states but can not do so because of the surrounding neighbors (\emph{hostile environment}), they get a \emph{booster} to upgrade their current status through $g$. This $g$ helps them achieve their desired status even if they have less number of neighboring cells to their \emph{support} (as $x\le K$). But whether the cell will be updated or not, is dependent on the probability value. Both cells with state $0$ and $1$ get this advantage uniformly in terms of the two probability distribution functions $\phi(x)$ and $\psi(x)$. As $g$ gives precedence towards some cells, it is to be applied with a caution -- so there is the \emph{upgrade} probability value $p$ which works as a controlling measure. Therefore, when $K=4$, $f$ works as a simple majority rule and depending on $g$ the system can be inclined towards a specific state. 

Note that, the parameters give us flexibility to design the model according to the need of an application. For example, for the model to have affection to converge to all-$0$ as a fixed point, we can set $\phi(x)$ and $\psi(x)$ accordingly. Similarly, we can change value of our parameter(s) to get different versions of the model which can be used for a specific purpose. In fact, we may also consider that in our model rule $g$ is applied with probability $p$ whereas the rule $f$ is applied with probability $(1-p)$ with $p$ being any probability value. This way of looking at these rules makes both of them \emph{temporally stochastic}. 
The next section shows some simulation results of our model to solve the affinity classification problem taking some specific value of the parameters.

%

\section{Solving Affinity Classification Problem: A Simulation}\label{simulation} 
\noindent We now simulate our proposed model to understand its efficacy in solving the affinity classification problem. As mentioned before, the model is a 2-dimensional finite CA that uses periodic boundary condition. For the simulation purpose, we consider here the grid size as $10^3\times10^3$, that is total number of cells = $10^6$. Further, our model is characterized by four parameters -- $K$, $\phi(x)$, $\psi(x)$ and $p$. In our simulation, we have assumed the following values for the parameters:
\begin{align*}
K&=4\\
\phi(x)&=\begin{cases}
0  &\text{if }  x \le 1\\
log_K(x) &{2\le x \le K }
\end{cases}\\
\psi(x)&=\begin{cases}
0  &\text{if }  x = 0\\
e^{x-K} &{1\le x \le K }
\end{cases}\\
p&=0.2
\end{align*}

\begin{figure}
	\centering
	\subfigure[ \label{fig:12}]{%
		\includegraphics[scale = 0.25]{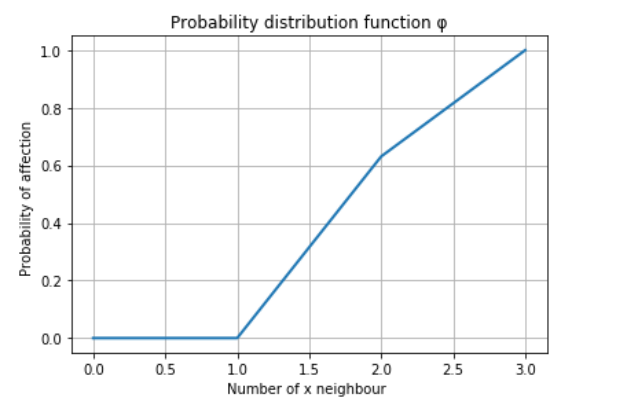}
	}
	\subfigure[ \label{fig:13}]{%
		\includegraphics[scale = 0.25]{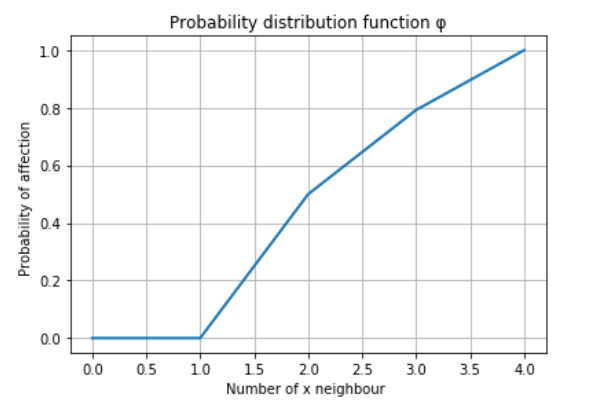}
	}
	\subfigure[ \label{fig:14}]{%
		\includegraphics[scale = 0.25]{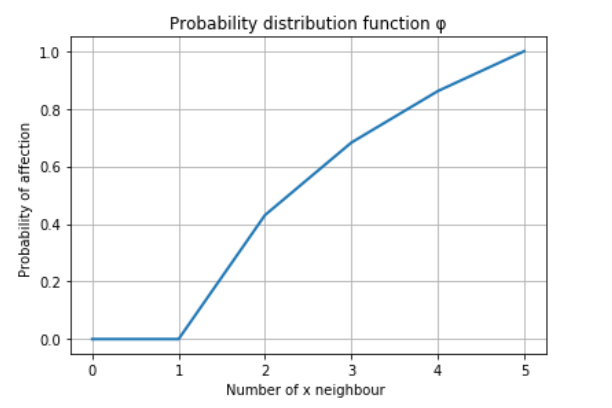}
	}
	\subfigure[ \label{fig:15}]{%
		\includegraphics[scale = 0.25]{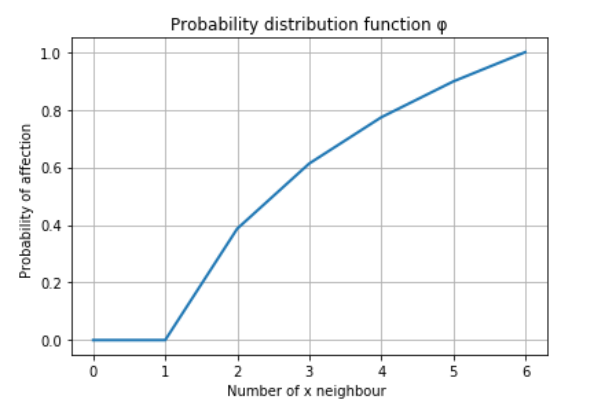}
	}     

	\caption{Graph of $\phi(x)$ for different $K$: a) $K=3$ ; b) $K=4$ ; c) $K=5$ ; d) $K=6$}
	\label{fig:16}
	
\end{figure}

\begin{figure}\centering
	\subfigure[ \label{fig:23}]{%
		\includegraphics[scale = 0.25]{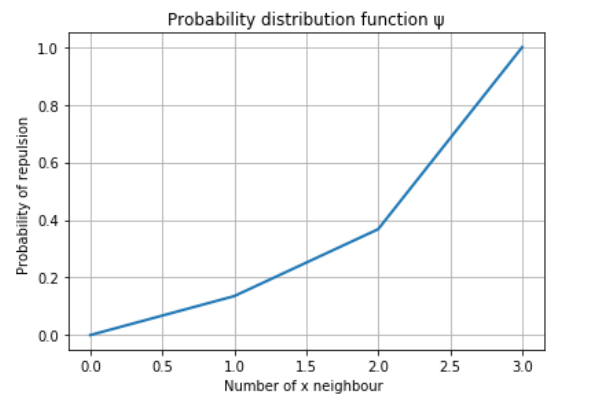}
	}
	\subfigure[ \label{fig:24}]{%
		\includegraphics[scale = 0.25]{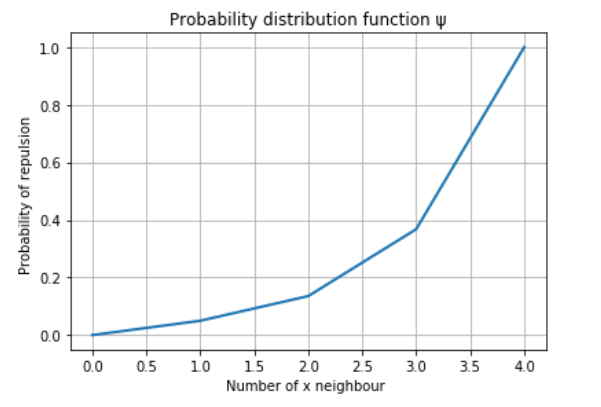}
	}
	\subfigure[ \label{fig:25}]{%
		\includegraphics[scale = 0.25]{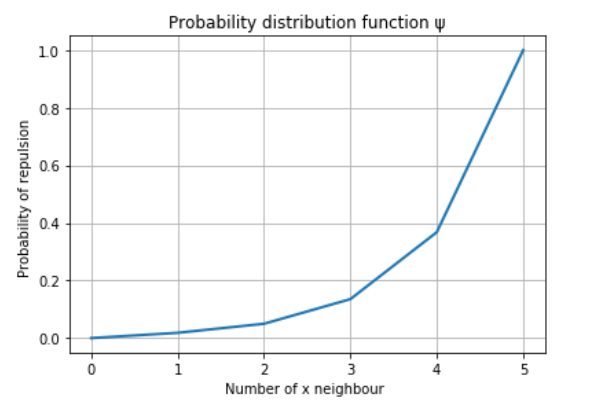}
	}
	\subfigure[ \label{fig:26}]{%
		\includegraphics[scale = 0.25]{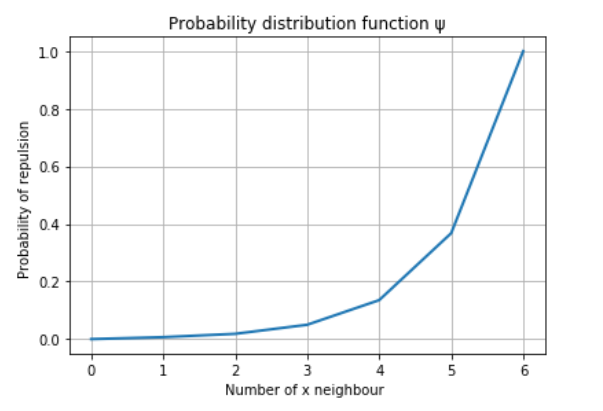}
	}     

	\caption{Graph of $\psi(x)$ for different $K$: a) $K=3$ ; b) $K=4$ ; c) $K=5$ ; d) $K=6$}
	\label{fig:27}
	
\end{figure}

As our model uses Moore neighborhood dependency on 2-D grid, $K$ is very small ($0\le K \le 8$). In this small range of $K$, logarithmic function grows faster than exponential function. Therefore, since we want to observe the affinity of the model towards all-$0$ configuration, we take $\phi(x)$ as a logarithmic function and $\psi(x)$ as an exponential function. As per our model, we use $x=0,1,\cdots,K$ to get the probability values for $\phi(x)$ and $\psi(x)$. We have plotted $\phi(x)$ and $\psi(x)$ for different $x$ to see their behavior at $K=4$ (see Figure \ref{fig:13} and Figure \ref{fig:24} respectively). We can observe that, at $K=4$, $\phi(1)=0.0$, $\phi(2)=0.5$, $\phi(3)=0.79248$, $\phi(4)=1.0$, whereas, $\psi(1)=0.0497$, $\psi(2)=0.1353$, $\psi(3)=0.3679$, $\psi(4)=1.0$. 
%
We observe that, if the count of supporting neighbors $x$ is increased then the probability of changing state from $1$ to $0$ is also increased (Figure~\ref{fig:13}); but, if $x$ is decreased then the probability of changing state from $0$ to $1$ is increased with the growth of the first function being faster than the latter (Figure~\ref{fig:24}).


\subsection{Random Initial Configuration} 
We have experimented our model with huge number of random initial configurations having various $\rho$ where 
$$\rho= \frac{\text{Number of 1s}}{\text{Total number of cells}} $$ 
Following are some sample results from our experiment when $K=4$. Here, $0$ is represented in color \emph{yellow} and $1$ is represented by color \emph{red}.

\begin{figure}\centering
	\subfigure[ \label{fig:28}]{%
		\includegraphics[scale = 0.20]{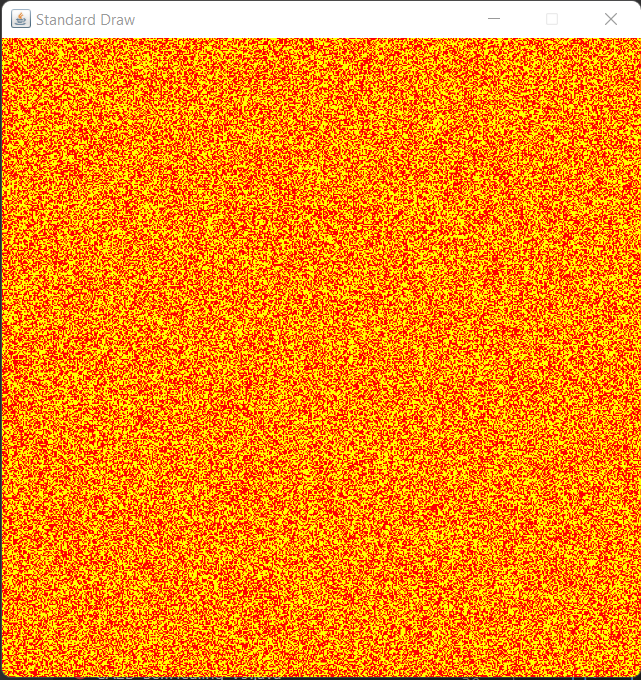}
	}
	\hfill
	\subfigure[ \label{fig:29}]{%
		\includegraphics[scale = 0.20]{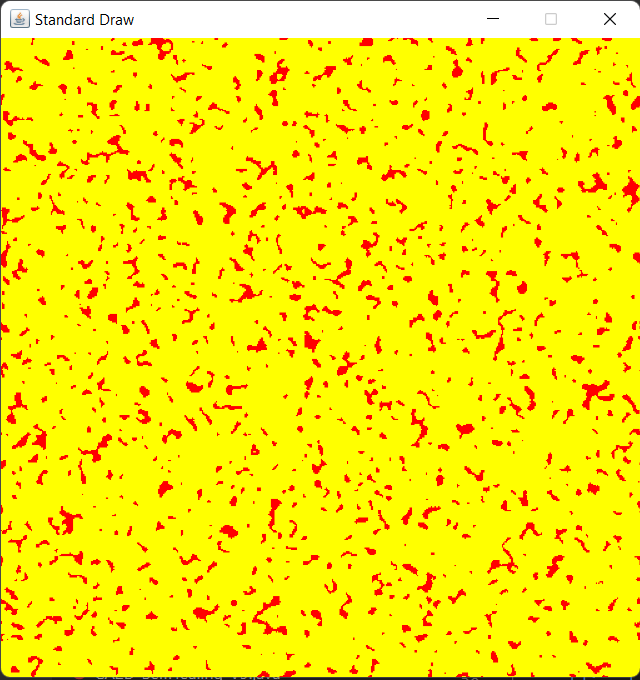}
	}
	\hfill  
	\subfigure[ \label{fig:30}]{%
		\includegraphics[scale = 0.20]{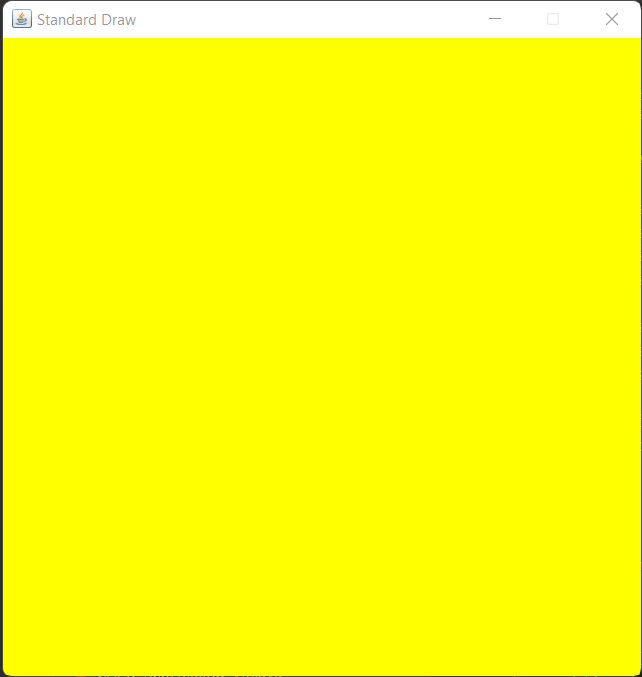}
	}
	
	\caption{For $K=4$ and $\rho=0.475$, the model converge after $150$ iterations: (a) Initial configuration, (b) An intermediate Configuration, (c)Final configuration(all-0)}
	\label{fig:31}
\end{figure}
\noindent Figure~\ref{fig:31} shows that, at $\rho=0.475$, for a random initial configuration, all the cells become yellow after $150$ iterations, that means, the model converge to it's converging point (all-$0$). We have experimented with large number of random initial configurations and seen that, in our experiments, when the initial configuration has $\rho \le 0.675$, the model is converging to all-$0$, otherwise it converges to all-$1$.
\begin{figure}\centering
	\subfigure[ \label{fig:41}]{%
		\includegraphics[scale = 0.20]{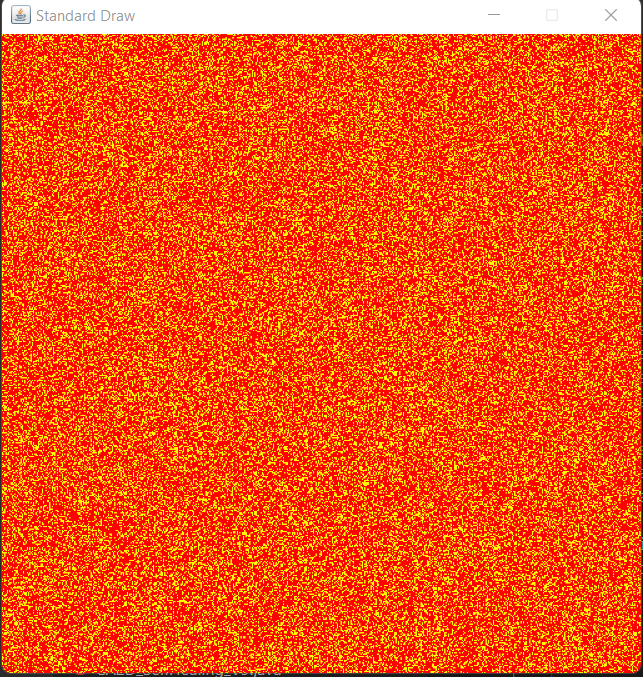}
	}
	\hfill
	\subfigure[ \label{fig:42}]{%
		\includegraphics[scale = 0.20]{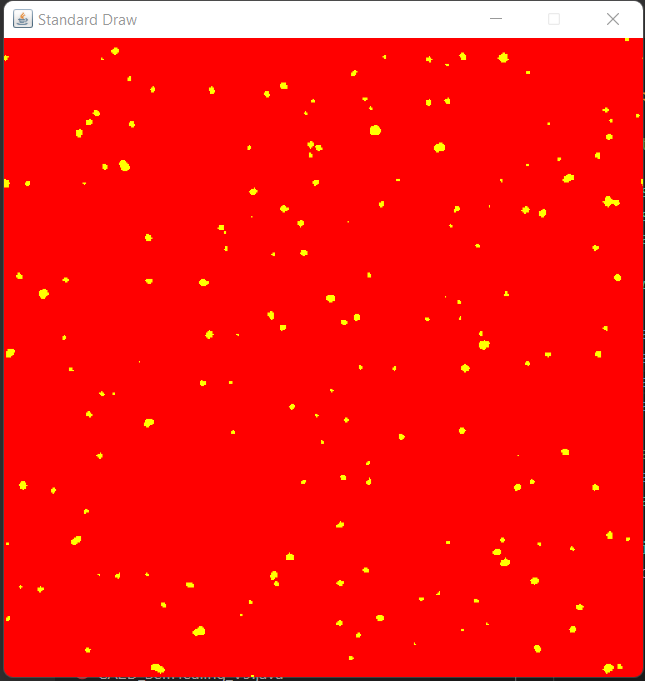}
	}
	\hfill  
	\subfigure[ \label{fig:43}]{%
		\includegraphics[scale = 0.20]{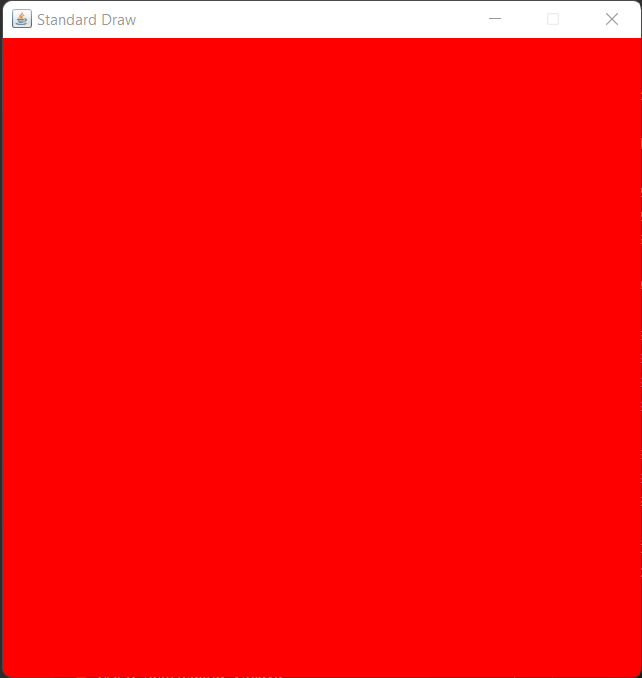}
	}
	
	\caption{For $K=4$ and $\rho=0.6989$: the model converge to all-$1$ after $137$ iterations: (a) Initial configuration, (b)An intermediate Configuration, (c)Final configuration(all-1)}
	\label{fig:4}
\end{figure}

Figure \ref{fig:4} shows another sample random initial configuration with an arbitrary $\rho > 0.675$ (here $\rho = 0.6989$). Here, the model converges to all-$1$ after 137 iterations.
\begin{figure}\centering
	\subfigure[ \label{fig:51}]{%
		\includegraphics[scale = 0.20]{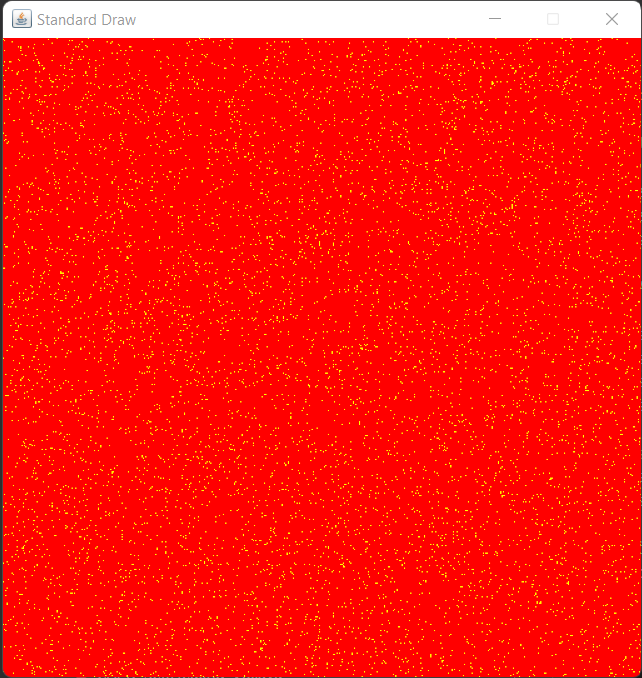}
	}
	\hfill
	\subfigure[ \label{fig:52}]{%
		\includegraphics[scale = 0.20]{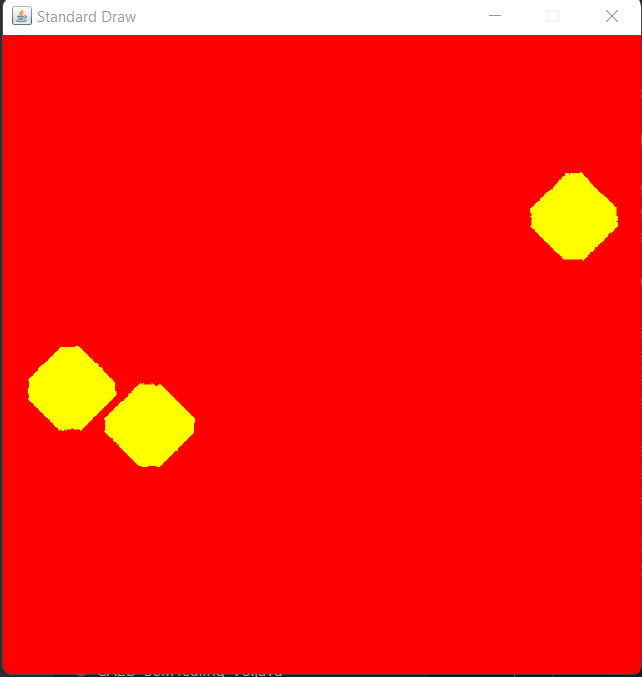}
	}
	\hfill  
	\subfigure[ \label{fig:53}]{%
		\includegraphics[scale = 0.20]{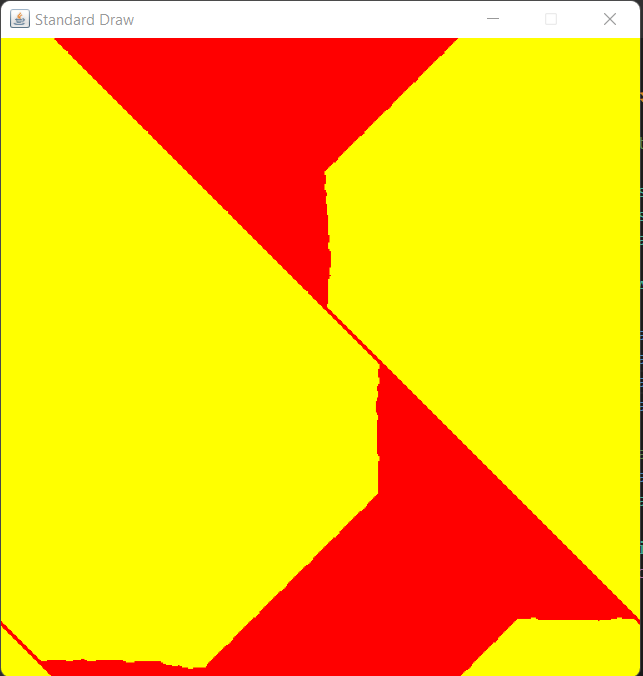}
	}
	\hfill  
	\subfigure[ \label{fig:54}]{%
		\includegraphics[scale = 0.20]{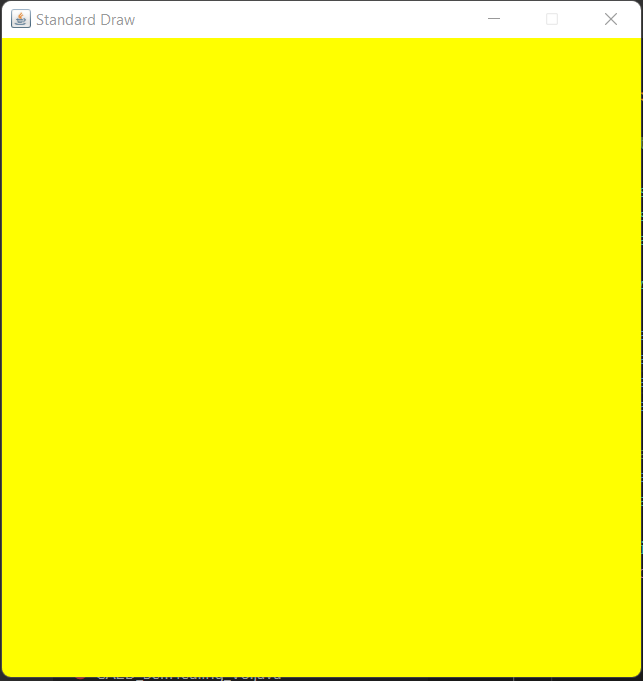}
	}
	
	\caption{For $K=3$ and $\rho=0.969852$, after 3132 iterations, the model converge to all-$0$'s: (a) Initial configuration, (b) An intermediate Configuration, (c) Another intermediate configuration, (d) Final configuration (all-$0$)}
	\label{fig:5}
\end{figure}

Naturally, the question comes, \emph{``Can we increase the affection probability so that even if we take a lot of $1$'s in the initial configuration then also the model converges to all-$0$?''}. To search for this answer, we have again done large number of experiments by varying the value of $K$. In our experiments, we have observed that when we decrease the value of $K$, the model is converging to all-$0$ even though $\rho > 0.68$. For example, for the initial configuration of Figure~\ref{fig:5}, if $K=3$, then although $\rho \le 0.96$, the model is converged to all-$0$s. By further experimentation, we observe that, if the value of $K$ is decreased to $2$ then $\rho$ can be as high as $0.99$ but the model may still converges to all-$0$. Similarly, when we increase the value of $K$ then the value of $\rho$ is to be decreased for converging to all-$0$. 

\begin{table}
	\centering
	\caption{Relationship between the values of $K$ and $\rho$ where the model converges to all-$0$ or all-$1$}\label{tab1}
	\resizebox{0.99\textwidth}{7.5cm}{
		\begin{tabular}{cc}
			\begin{tabular}{|c|c|c|c|}
				\hline\hline
				{\bfseries K } & {\bfseries $\rho$} & {\bfseries  Number of iterations } & {\bfseries Converge to }\\
				\hline
				1 &  0.000002 & 1 &all-0\\
				1 &  0.0002 & 2 &all-0\\
				1 &  0.0051 & 2 &all-0\\
				1 &  0.3 & 3 &all-0\\
				1 &  0.55 & 4 &all-0\\
				1 &  0.67 & 6 &all-0\\
				1 &  0.6864 & 8 &all-0\\
				1 &  0.943 & 21 &all-0\\
				1 &  0.991 & 76 &all-0\\
				1 &  0.997 & 170 &all-0\\
				1 &  0.9995 & 489 &all-0\\
				\hline
				2 &  0.1 & 3 &all-0\\
				2 &  0.3 & 4 &all-0\\
				2 &  0.4 & 5 &all-0\\
				2 &  0.61 & 7 &all-0\\
				2 &  0.74 & 14 &all-0\\
				2 &  0.8 & 16 &all-0\\
				2 &  0.9536 & 90 &all-0\\
				2 &  0.965 & 151 &all-0\\
				2 &  0.982 & 323 &all-0\\
				2 &  0.993 & 759 &all-0\\
				2 &  0.995 & 3 &all-1\\
				2 &  0.9995 & 2 &all-1\\
				\hline
				3 &  0.1 & 3 &all-0\\
				3 &  0.3 & 6 &all-0\\
				3 &  0.4 & 9 &all-0\\
				3 &  0.55 & 20 &all-0\\
				3 &  0.61 & 22 &all-0\\
				3 &  0.6864 & 44 &all-0\\
				3 &  0.8 & 109&all-0\\
				3 &  0.943 & 518 &all-0\\
				3 &  0.953 & 2042 &all-0\\
				3 &  0.96 & 2372 &all-0\\
				3 &  0.965 & 3220 &all-0\\
				3 &  0.982 & 3 &all-1\\
				3 &  0.991 & 3 &all-1\\
				3 &  0.995 & 2 &all-1\\
				
				\hline\hline
			\end{tabular}
			&
			\begin{tabular}{|c|c|c|c|}
				\hline\hline
				{\bfseries K } & {\bfseries $\rho$} & {\bfseries  Number of iterations } & {\bfseries Converge to }\\
				\hline
				
				4 &  0.1 & 4 &all-0\\
				4 &  0.4 & 66 &all-0\\
				4 &  0.55 & 805 &all-0\\
				4 &  0.61 & 3074 &all-0\\
				4 &  0.65 & 7019 &all-0\\
				4 &  0.67 & 12385 &all-0\\
				4 &  0.675 & 16186 &all-0\\
				4 &  0.6864 & 261 &all-1\\
				4 &  0.7 & 159 &all-1\\
				4 &  0.74 & 69 &all-1\\
				4 &  0.8 & 8 &all-1\\
				4 &  0.943 & 4 &all-1\\
				4 &  0.965 & 4 &all-1\\
				4 &  0.991 & 2 &all-1\\
				
				\hline
				5 &  0.06 & 8 &all-0\\
				5 &  0.08 & 14 &all-0\\
				5 &  0.09512 & 8 &all-0\\
				5 &  0.1  & 14 &all-0\\
				5 &  0.3 & 304 &all-1\\
				5 &  0.4  & 91 &all-1\\
				5 &  0.55 & 12 &all-1\\
				5 &  0.61  & 10 &all-1\\
				5 &  0.686 & 6 &all-1\\
				\hline
				6 &  0.001 & 2 &all-0\\
				6 &  0.0051 & 3 &all-0\\
				6 &  0.00994 & 8 &all-0\\
				6 &  0.03 & 638&all-1\\
				6 &  0.0629 & 230 &all-1\\
				6 &  0.076 & 194 &all-1\\
				6 &  0.08 & 98 &all-1\\
				6 &  0.1 & 58 &all-1\\
				\hline
				7 &  0.000002 & 2 &all-0\\
				7 &  0.0002 & 2 &all-0\\
				7 &  0.0004 & 2 &all-0\\
				7 &  0.0005 & 976 &all-1\\
				7 &  0.0009 & 375 &all-1\\
				7 &  0.001 & 417 &all-1\\
				7 &  0.00499 & 136 &all-1\\
				7 &  0.00994 & 83 &all-1\\
				7 &  0.0676 & 25 &all-1\\
				7 &  0.1 & 12 &all-1\\
				7 &  0.55 & 4 &all-1\\
				7 &  0.95 & 2 &all-1\\
				\hline\hline
			\end{tabular}
	\end{tabular}}
\end{table}

Figure \ref{fig:16}and \ref{fig:27} show the variation of the probability distribution functions for different $K$ values. {If $K$ is changed then the growth of the probability distribution functions $\phi(x)$ and $\psi(x)$ are also changed with respect to $K$.} Table~\ref{tab1} gives some of our experimental results. In each of the subtables of this table, column $1$ and $2$ describe the initial configurations in the form of $K$ and $\rho$, whereas, column $3$ and $4$ show experimental outcomes.

\begin{table}
	\centering
	\caption{Relationship between the values of $K$ and the block of $0$s where the model converges to all-$0$ or all-$1$ considering number of $0$s are placed sequentially in the grid}\label{tab2}
	\resizebox{0.6\textwidth}{!}{
		\begin{tabular}{|c|c|c|c|}
			\hline\hline
			{\bfseries $K$} & {\bfseries Number of $0s$ } & {\bfseries Iterations (Time steps) }& {\bfseries Converges to }\\
			\hline
			1 &  1 & 1 &all-1\\
			1 &  2 & 998 &all-0\\
			\hline
			2 &  2 & 1 &all-1\\
			2 &  3 & 1152 &all-0\\
			2 &  4 & 1148 &all-0\\
			\hline
			3 &  3 & 1 &all-1\\
			3 &  4 & 3874 & all-0\\
			3 &  25 & 4093 & all-0\\
			\hline
			4 &  49 & 64 &all-1\\
			4 &  64 & 92 &all-1\\
			4 &  70 & 492 &all-1\\
			4 &  81 & 576 &all-1\\
			4 &  100 & 15915 &all-0\\
			4 &  144 & 16138 &all-0\\
			\hline\hline
	\end{tabular}}
\end{table}

\subsection{Initial Configuration with Block of $0$s and $1$s}\label{sec:block}
Previous subsection shows the results when $0$ and $1$ in the initial configuration are randomly organized. Now, we experiment with initial configurations where block of cells are set to have same value. Table~\ref{tab2} and \ref{tab3} depict our sample results. In Table \ref{tab2}, we consider initial configurations with a small number of consecutive cells at state $0$ and the remaining cells in state $1$. For every value of $K$ ($ 1 \le K \le 7$), column 2 shows the number of consecutive cells having same value in our experiments so that the model converges to all-$0$ . For instance, when $K=4$ then an initial configuration having a block of $100$ consecutive $0$s converges the model to all-$0$. These consecutive $0$s form a cluster and it grows in size to converge the model to all-$0$.
\begin{figure}\centering
	\subfigure[ \label{fig:17}]{%
		\includegraphics[scale = 0.20]{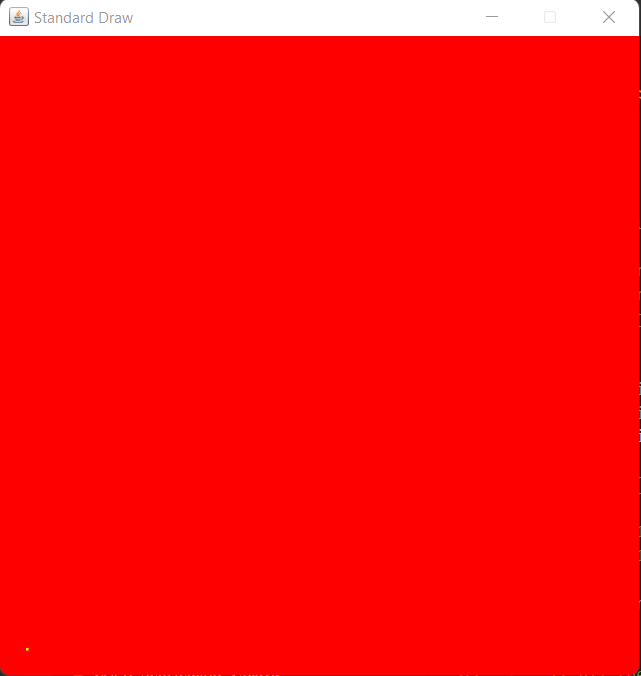}
	}
	\hfill
	\subfigure[ \label{fig:18}]{%
		\includegraphics[scale = 0.2]{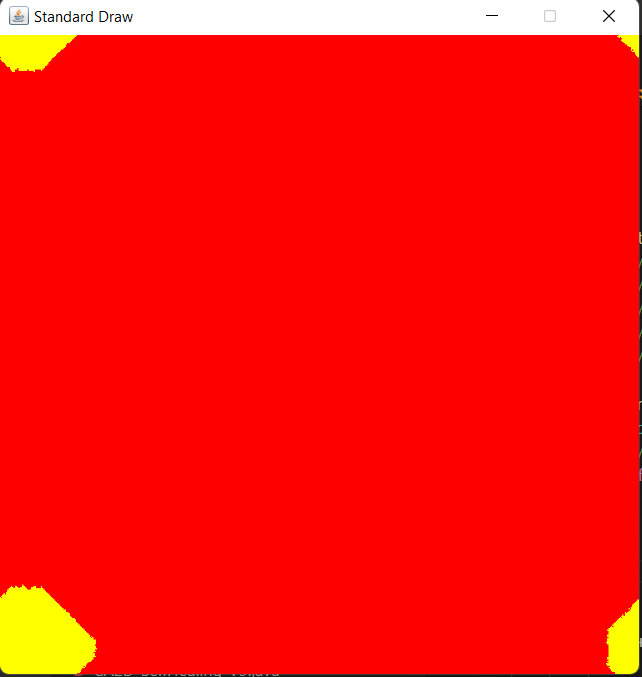}
	}
	\hfill  
	\subfigure[ \label{fig:19}]{%
		\includegraphics[scale = 0.2]{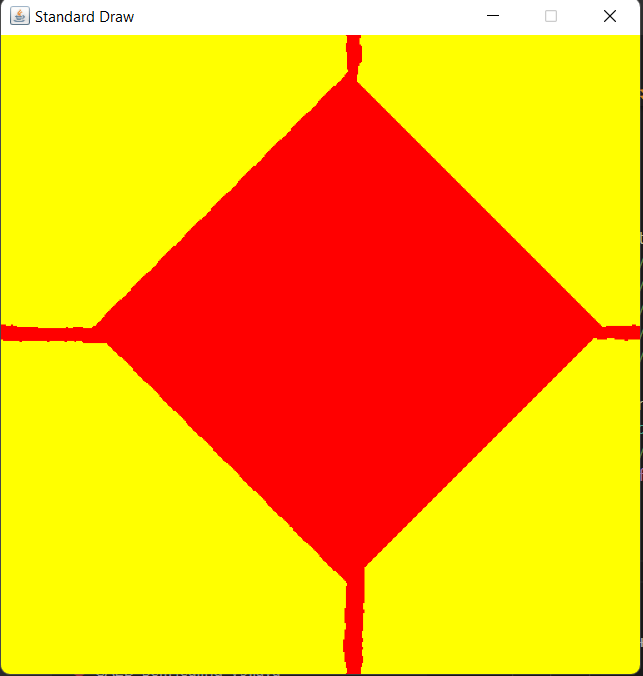}
	}
	\hfill
	\subfigure[ \label{fig:20}]{%
		\includegraphics[scale = 0.2]{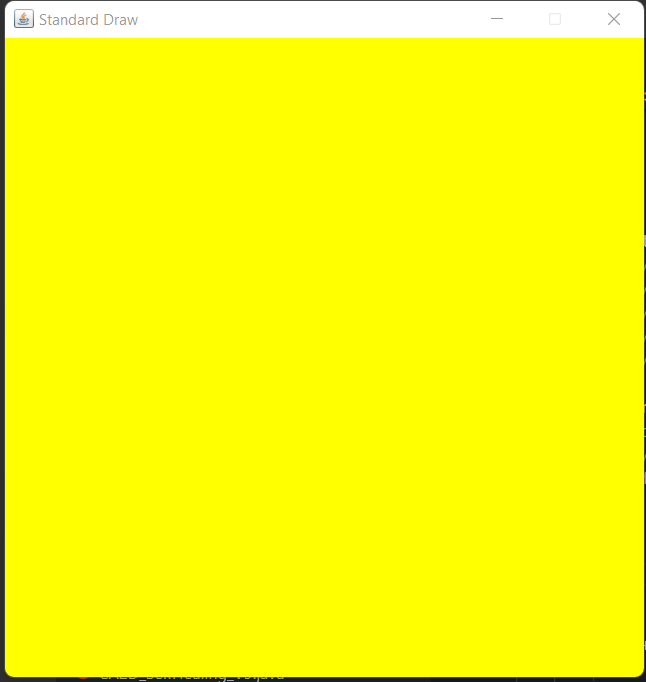}
	}     

	\caption{$K=3$ and the block of $25$ $0s$, the model converge to all $0s$ after 4093 iterations: a) Initial configuration, b)Intermediate Configuration, c)Another intermediate configuration, d) Final Configuration converge to all-$0$}
	\label{fig:21}
	
\end{figure}
Figure~\ref{fig:21} shows a random initial configuration with $10^6$ cells, where only $25$ consecutive cells are in state $0$ and the value of $K$ is $3$. We can observe that, although the number of $0s$ is very less, still the model converges to all-$0$ after $4093$ iterations (see Figure~\ref{fig:20}). {Therefore, our model has affinity to converge to all-$0$ even if at initial configuration the number of $0$s is very less in comparison to the grid size.} 

However, the model does not always converge to the desired fixed point. For example, at $K=3$, for a random initial configuration having block of $0$s of size $81$, the model converges to all-$1$ (Table~\ref{tab2}).
\begin{table}
	\centering
	\caption{Relationship between the values of $K$ and the block of $0s$ where the model converges to all-$0$ or all-$1$ considering number of $1s$ are placed sequentially in the grid}\label{tab3}
	\resizebox{0.6\textwidth}{!}{
		\begin{tabular}{|l|l|l|l|}
			\hline\hline
			{\bfseries $K$ } & {\bfseries Number of $1$'s }& {\bfseries Iterations (Time steps) } & {\bfseries Attractor }\\
			\hline
			5 &  25 & 24 &all-0\\
			5 &  225 & 212 &all-0\\
			5 &  256 & 483 &all-0\\
			5 &  324 & 12148 &all-1\\
			5 &  400 & 11870 &all-1\\
			\hline
			6 &  2 & 1 &all-0\\
			6 &  4 & 13 &all-0\\
			6 &  5 & 1519 &all-1\\
			6 &  6 & 1510 &all-1\\
			6 &  8 & 1507 &all-1\\
			6 &  9 & 1493 &all-1\\
			
			\hline
			7 &  1 & 1 &all-0\\
			7 &  2 & 979 &all-1\\
			
			\hline\hline
			
	\end{tabular}}
\end{table}
Table \ref{tab3} depicts some sample results from our experiment where we take small number of $1$s organized in sequential order that is, they make cluster. Here, we can see that, after taking the $K>4$, sometimes the model converges to all-$1$s even the number of $1$'s is very less. 
Therefore, even if the model has an affection to converge to all-$0$, the value of $K$ may take a major role to converge the model in a direction (all-$0$ or all-$1$).

\section{Applications}\label{application}
\noindent As discussed in Section~\ref{model}, the parameters give us flexibility to design our model according to the need of the solution to a particular problem. There are several possible applications of our model. Here we discuss some of them.

\subsection{Modeling Self-healing Systems}\label{sec:application}
Living systems are assumed to be more intelligent than a non-living system. Therefore, to be intelligent, a machine (non-living system) has to emulate the properties of living systems. Among the properties, self-healing is a basic and important biological property which indicates sign of life. 
Self-healing is the ability to reorganize and heal itself. 
If a machine has self-healing ability, it is likely to mimic other properties of living elements like self-replication. Hence, it will be more intelligent just like a living system. We can show that our proposed CA can be used to model any self-healing system where parameters of our abstract model can be interpreted as the characteristics of the self-healing system.

Let us interpret our model as the following. Let the grid of cells embodies a collection of living elements (they can be cells, humans, animals -- anything), where state $0$ means the cell is healthy and $1$ means it is sick. We want to model how much infection the cells can endure and still heal. 
By default, the living system is healthy, that is, all cells are in state $0$. Now, suppose, because of some change in environment, a number of cells get infected and update their states to $1$ (become sick). This is our initial configuration in the model. We start to observe the dynamics of the system from here. Let us consider that, in our model, system's immunity is the immunity of individual cells and as a whole the system's \emph{health} is the \emph{majority} of the \emph{individual} cell's health condition. So, at the initial configuration, if we ask the system, ``\emph{Are you sick?}'', it can answer ``Yes'' or ``No'' depending on density of $1$ ($\rho$). If using this model the system can \emph{heal} itself, that is, comes back to all-$0$, then we can call the model as a model for self-healing systems. At that time, the answer to ``\emph{Are you sick?}'' to the model will always be ``No''. Therefore, our target is to converge the grid to all-$0$ so that we can say there is no infection and ``\emph{The model is {Not Sick}}''. However, if the model converges to all-$1$ then we have to declare, ``\emph{The model is {Sick}}''.

Now, any living body has some inbuilt immunity status. This immunity is represented by the first parameter $K$. Just like immunity is different for different elements, $K$ itself is a variable. When $K=4$, the system can be interpreted as the situation of natural immunity having no prevailing sickness. The deterministic rule $f$ plays the role of natural healing process based on immunity $K$. Results from Section\ref{simulation} show that, if converging point is set to all-0 and $K\le4$ then there is a tendency to converge towards all-$0$ even if in the initial configuration number of 1s $>$ number of $0$s. This indicates, like any living body, our model also wants to become \textit{Not Sick}. 
%
%



\begin{figure}\centering
	\subfigure[ \label{fig:h3}]{%
		\includegraphics[scale = 0.2]{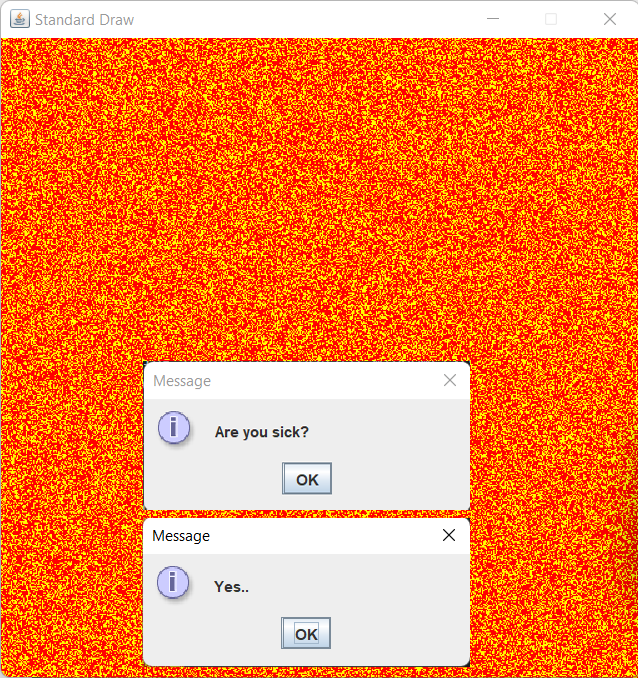}
	}
	\hfill
	\subfigure[ \label{fig:h4}]{%
		\includegraphics[scale = 0.2]{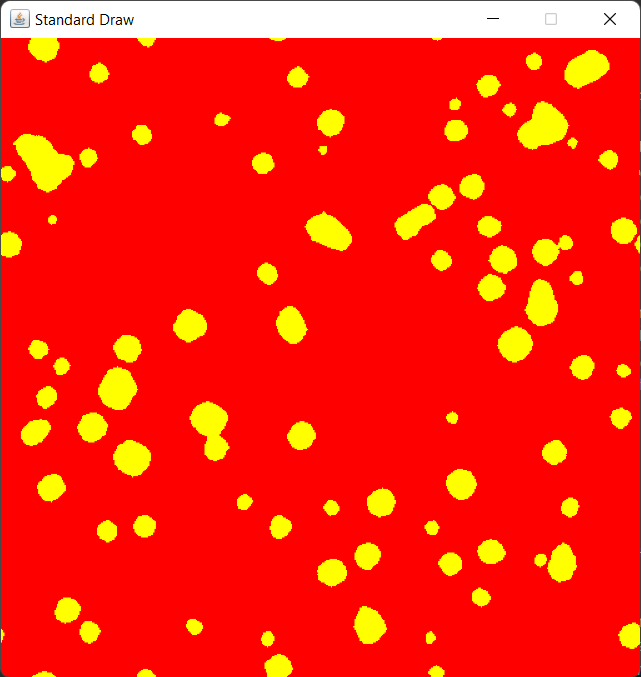}
	}
	\hfill  
	\subfigure[ \label{fig:h5}]{%
		\includegraphics[scale = 0.2]{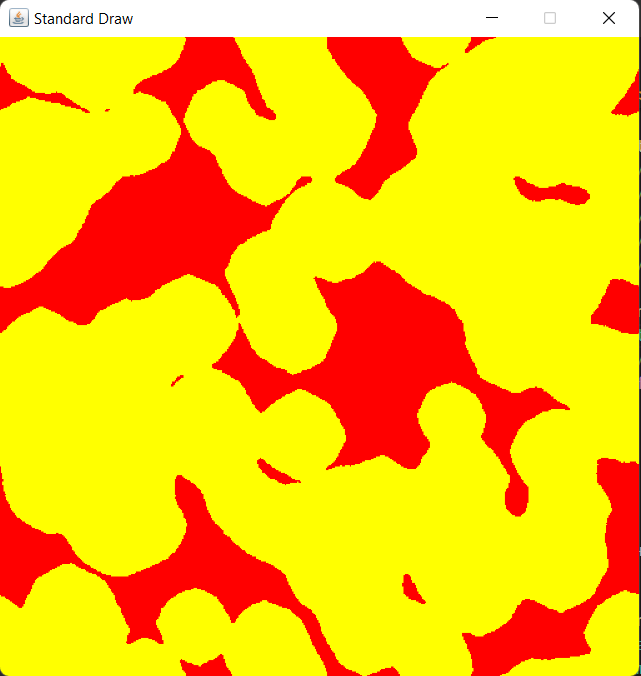}
	}
	\hfill
	\subfigure[ \label{fig:h6}]{%
		\includegraphics[scale = 0.2]{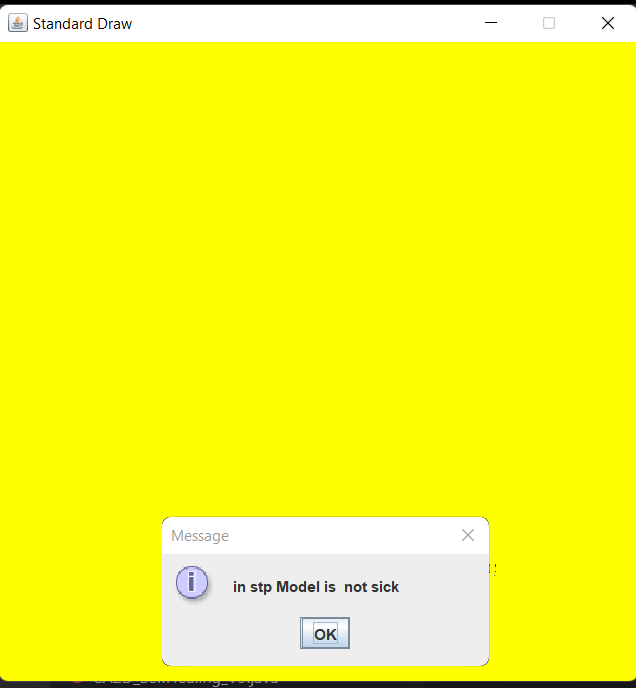}
	}     

	\caption{(a)Initial configuration of a sick model ; (b) and (c) shows two intermediate configurations during evaluation ; (d) The model is healed}
	\label{fig:h7}
	
\end{figure}
However, even in this condition, if number of infected cells become too large ($\rho$ is high), then, according to our rule, the system is \emph{sick}.  So, the inherent immunity is not enough to restore it to its health. For example, if we take a random initial configuration with some infected cells (cell state $1$) where $\rho=0.632275$ ($\rho$ = density of $1$'s) and $K=4$, then, at this stage the model is \textit{Sick} (see Figure \ref{fig:h3}). At this point, the cells are given some \emph{booster} to improve its immunity in terms of $g$. Here, $g$ may be considered as a \emph{vaccine} for the infection. As if, it can bypass the \emph{natural justice} process giving the cells a second chance to live. 
But, whether vaccine will be effective to a cell, is not deterministic (so, $g$ is stochastic). Further, when this vaccine is to be applied to the system is also not pre-determined (temporally stochastic CA with probability $p$). 

Nevertheless, for every cell, the vaccine will not react similarly. A large number of sick cells with favorable environment may become healthy ($\phi(x)$), whereas, some healthy cells with unhealthy environment can become sick ($\psi(x)$). But, if we take $\phi(x)$ as logarithmic and $\psi(x)$) exponential like defined in Section~\ref{simulation}, by choosing $K$ and $x$, we can see that, after some iterations the model converge to all-$0$ (see Table~\ref{tab1}). Then we can say the model is \textit{Not Sick} (Figure~\ref{fig:h6}).

However, if we increase $\rho$ value further (say, from $0.675$ to $0.68$ or more), then for the same $K$ and $x$, the model may be converged to all-$1$ (see Table\ref{tab1}) and the model becomes \textit{Sick}. Therefore, the role of $K$ and $x$ is very important to model self-healing systems. If we want to have our system a larger tendency to heal, then we need to choose the parameters of our model wisely. Moreover, if the affection probability $\phi(x)$ is large, then the system has more tendency to heal.
This probability indicates the ability to repair or heal oneself automatically and evolve oneself according to the demand of the environment.
%

This is how the self-healing is modeled by our CA. It also shows that, our abstract model can be a good interpretation of the role of vaccination in living population. Also, observe that, our proposed model takes the global decision democratically where every single cell take their own decision and the system comes to a consensus. Because of these properties we claim that our proposed model is intelligent.


\subsection{Modeling Transformation Process}\label{sec:intelligence}
In nature and chemical world, we get glimpses of several transformation processes -- water evaporates into vapor, a drop of color in a glass of liquid dissolves giving the whole glass of liquid a lighter shade of that color. All these processes happen to conserve the law of mass and energy. This section shows that our CA can be used to model such transformation processes.

\begin{figure}\centering
	\subfigure[$t=0$ \label{fig:s1}]{%
		\includegraphics[scale = 0.15]{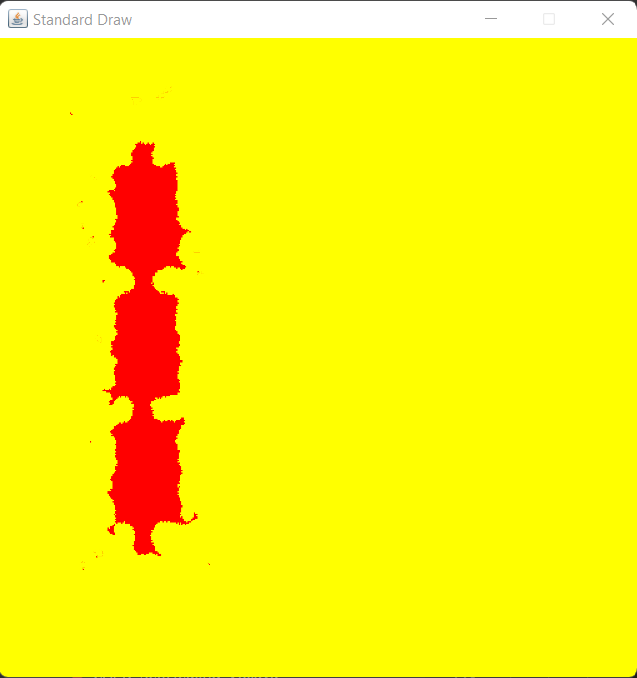} 
	}
	\hfill	
	\subfigure[$t=42$ \label{fig:s2}]{%
		\includegraphics[scale = 0.15]{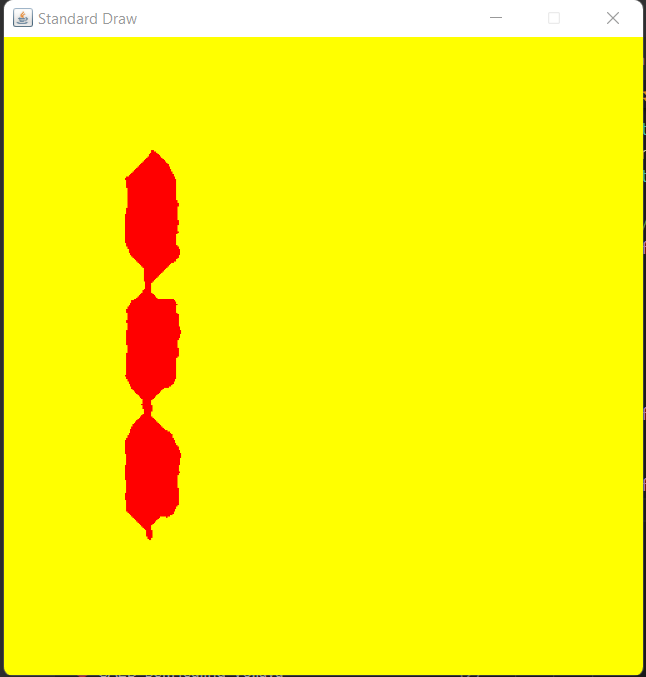} 
	}
	\hfill	
	\subfigure[ $t=78$ \label{fig:s3}]{%
		\includegraphics[scale = 0.15]{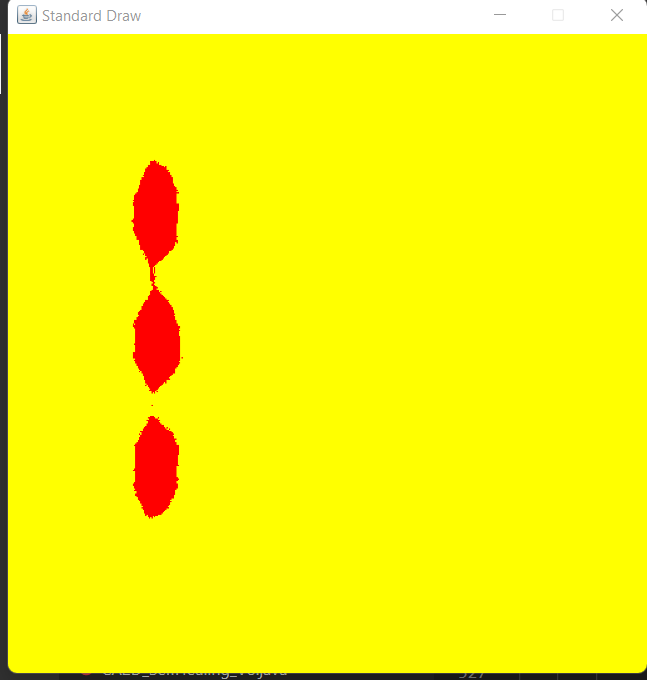}
	}
	\hfill
	\subfigure[$t=108$  \label{fig:s4}]{%
		\includegraphics[scale = 0.15]{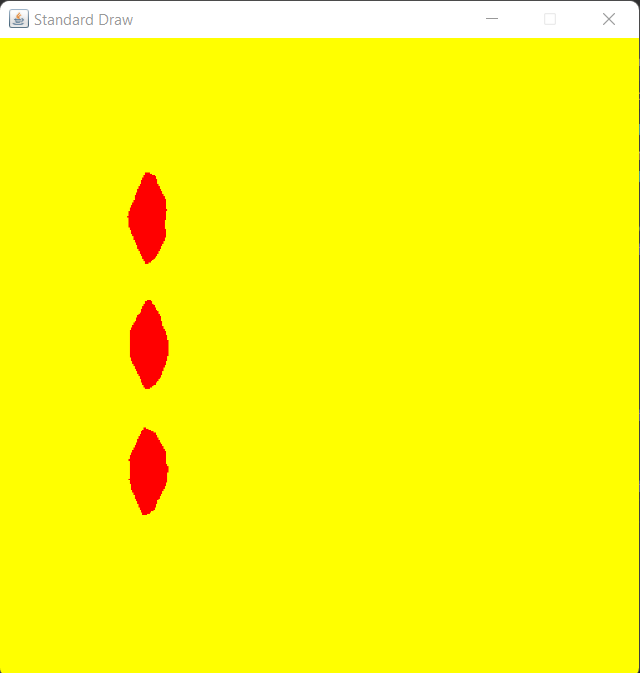}
	}
	\hfill  
	\subfigure[ $t=151$ \label{fig:s5}]{%
		\includegraphics[scale = 0.15]{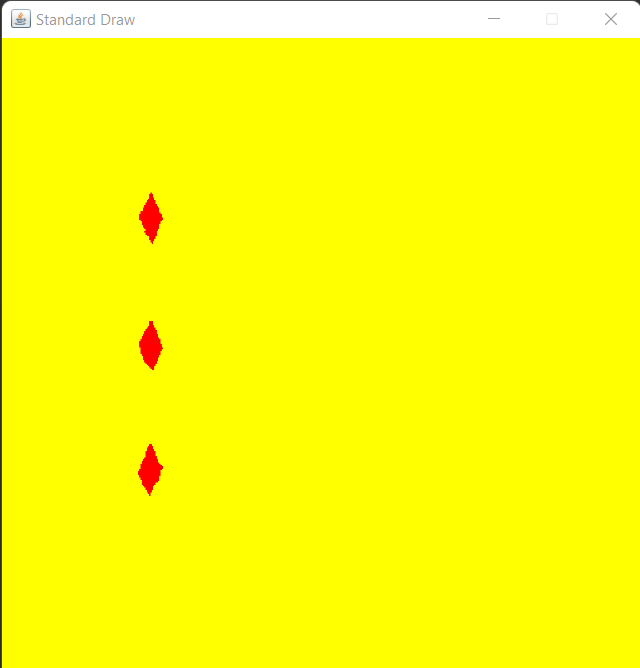}
	}
	\hfill
	\subfigure[$t=180$  \label{fig:s6}]{%
		\includegraphics[scale = 0.15]{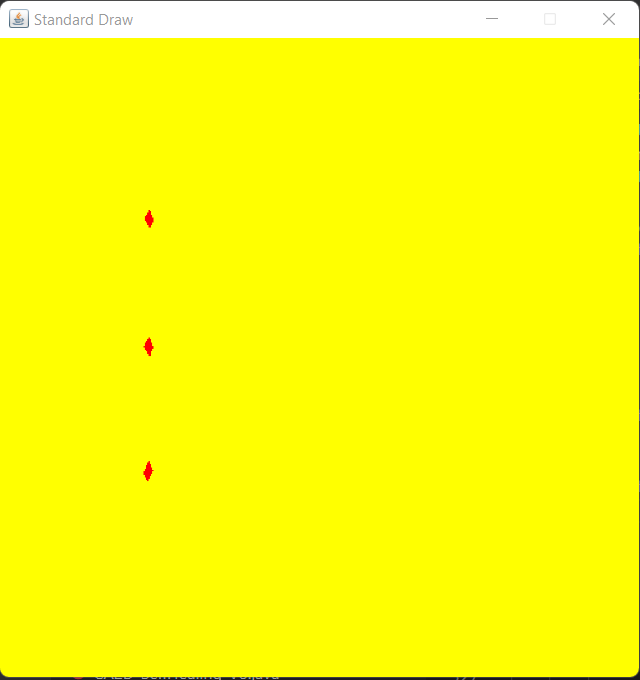}
	}     
	\hfill
	\subfigure[$t=198$  \label{fig:s7}]{%
		\includegraphics[scale = 0.15]{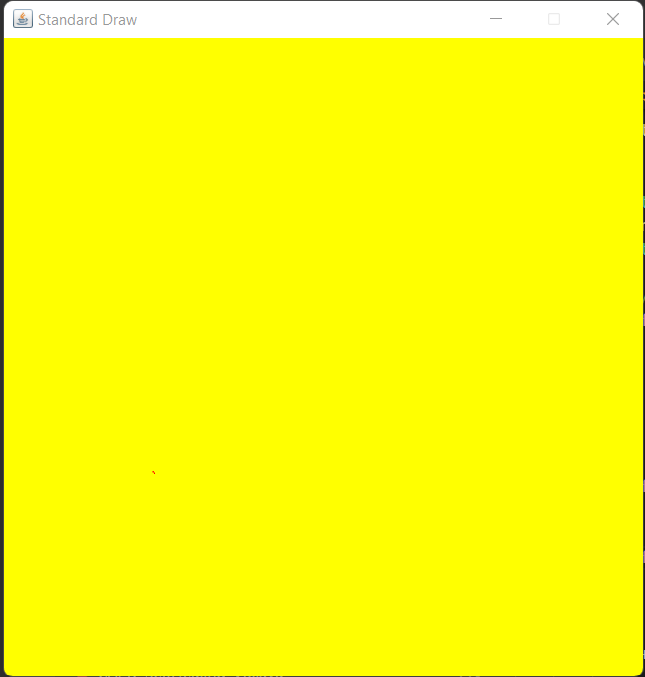}
	}    
	\caption{Simulation of a transformation process}
	\label{fig:7}
	
\end{figure}

During the process of transformation, the particles are divided into smaller sized particles and dissolves until the system comes to an equilibrium. In our model, if we set $K=3$ (and other parameters as same as Section~\ref{simulation}), then, for some special initial configurations, the evolution of the CA looks like transformation processes -- the configuration is divided into two or more smaller configurations. It goes on dividing and dissolving until the system converges to a fixed point which signifies the equilibrium state. For example, in Figure \ref{fig:7}, 
an initial configuration is shown, which, after some iterations, is divided into more than three configurations. It keeps on getting smaller until it converges to the fixed point all-$0$ when the system has reached its equilibrium. Hence, we can say that, by varying the parameters of our model, we can simulate the transformation process from one system to another by our CA.


\subsection{Density Classification Problem}
The \emph{density classification problem} can also be addressed by our model.
According to the definition of \emph{affinity classification problem}, this problem goes down to the former if we take the density of $1$s ($\rho$)=$0.5$. 
However, here, instead of taking $\rho$ as exact $0.5$, we take it as a variable and see how close we can reach to solve this classical problem using our model.

Previous works have established that density classification problem is not solvable by spatially stochastic CA (uniform or non-uniform), but can be solved by using temporally non-uniform CA \cite{Fuk05,fuks2015solving}. So, we also take our CA as temporarily non-uniform with a stochastic component $(g)$ which perfectly fits our model. However, a property of this problem is, there is no affinity towards any state at any time. Hence, to make the system unbiased, we take the number of neighbors required to change from one state to another $(K)$ as $4$. Also, we choose both the probability distribution functions $\phi(x)$ and $\psi(x)$ to be same. That is, if a cell is at state $1$ and it has $x$ number of neighbors with state $0$, it updates its value to $0$ with the same probability distribution function as in case of the cell being at state $0$ with $x$ number of neighbors with state $1$ and gets updated to state $1$. Further, we consider the upgrade probability value $p=0.1$ such that the stochastic component ($g$) is applied with very low probability. 

Here, we show simulation results for two different probability distribution functions -- linear and exponential. 
For the first case, the value of the parameters for the model are:
\begin{align*}
K&=4\\
\phi(x)&=\begin{cases}
\frac{x}{K} &\text{for }  0 \le x \le K\\
\end{cases}\\
\psi(x)&=\begin{cases}
\frac{x}{K} &\text{for }  0 \le x \le K\\
\end{cases}\\
p&=0.1
\end{align*}
We have done huge experimentation on random initial configurations over $200 \times 200$ grid based on this model. Some sample simulation results are shown in Table~\ref{tab4}. Here, the first column indicates some $\rho$ values whereas, the third and fourth columns represents that, for each of these $\rho$, out of $100$ experiments how many are converged to all-$0$ and all-$1$ respectively. {In our experiments, we observe that when initial configuration is taken randomly, then for $\rho \le 0.4647$ or $\rho \ge0.54$, our model converges to its fixed point (all-$0$ and all-$1$ respectively). 
	
	\begin{table}
		\caption{Taking 2D-square grid ($200 \times 200$) and both $\phi$ and $\psi$ as linear functions}\label{tab4}
		\centering
		\resizebox{0.9\textwidth}{!}{
			\begin{tabular}{|c|c|c|c|} \hline  \hline   
				$\rho$ (number of $1$'s) & Number of experiments & Converge to all - $0$ & Converge to all - $1$\\  \hline 
				$\le$ 0.4647 & 100&100&0 \\
				0.4779710& 100 & 96&4 \\
				0.49112875& 100 & 73&27 \\
				0.5036035& 100 & 37 & 63  \\
				0.51548925 & 100 & 9 & 91\\
				0.52758225 & 100 & 5 & 95 \\
				0.5394037& 100 & 1 & 99\\
				$\ge$ 0.54 & 100 & 0 & 100 \\
				\hline \hline 
		\end{tabular}}  
	\end{table} 
	
	For the second case, we take both $\phi$ and $\psi$ as exponential functions with $K=4$ and $p=0.1$. Hence, the changed parameters of the model are:
	\begin{align*}
	\phi(x)&=\begin{cases}
	0  &\text{if }  x =0\\
	e^{x-K}  &\text{for }  1 \le x \le K\\
	\end{cases}\\
	\psi(x)&=\begin{cases}
	0  &\text{if }  x =0\\
	e^{x-K}  &\text{for }  1 \le x \le K\\
	\end{cases}
	\end{align*}
	We again repeat our experiments with a large set of random initial configurations over $100 \times 100$ grid. Table~\ref{tab5} shows some sample results of this experiment.

	\begin{figure}\centering
		\vspace{-1.5em}
		\subfigure[\label{fig:10}]{%
			\includegraphics[scale = 0.2]{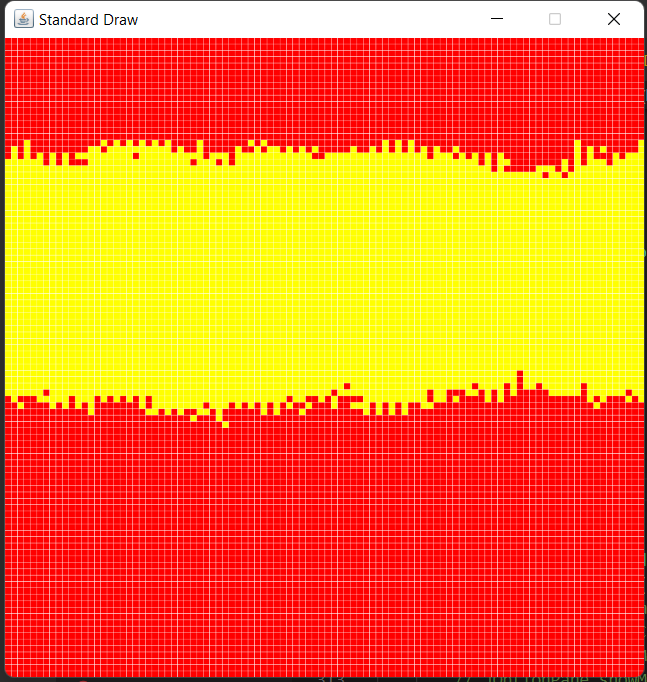}
		}
		\hfill
		\subfigure[\label{fig:11}]{%
			\includegraphics[scale = 0.2]{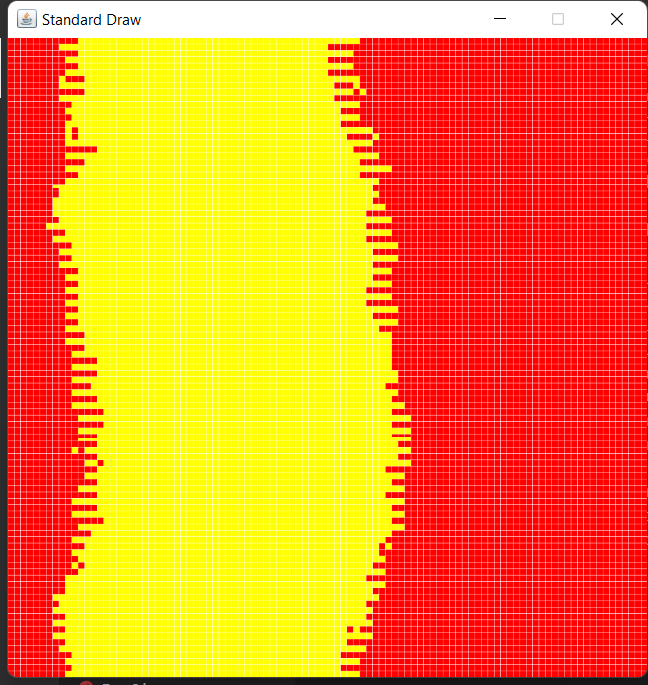}
		}    

		\caption{Some unsolvable configurations for density classification problem in 2D square grid}
		
	\end{figure}
	\begin{table}
	\caption{Taking 2D-square grid with size $100 \times 100$ and both $\phi$ and $\psi$ as exponential functions}\label{tab5}
	\centering
	\resizebox{0.9\textwidth}{!}{
		\begin{tabular}{|c|c|c|c|} \hline  \hline   
			$\rho$ (number of $1$'s) & Number of experiments & Converge to all - $0$ & Converge to all - $1$\\  \hline 
			$\le$ 0.4679  & 100 & 100 & 0  \\
			0.47837 & 100 & 96 & 4  \\
			0.513014 & 100 & 30 & 70  \\
			0.5181367 & 100 & 0 & 100  \\
			$\ge$ 0.520 & 100 & 0 & 100  \\
			\hline \hline 
	\end{tabular} }
\end{table} 

	Here also we observe that, when the initial configuration is random, then for $\rho \le 0.4679$ or $\ge 0.520$, the model reaches its desired fixed point (all-$0$ or all-$1$). However, when the configurations are block of $0$s or $1$s forming a cluster, then it fails to reach the desired fixed point. Figure~\ref{fig:10} shows example of two such patterns where the model can not reach its fixed point (see Section~\ref{sec:block} for more details). 
	\section{Conclusion and Future Scopes}
	\label{future}
	\noindent There are several properties in the living system that make them intelligent -- affection is one of them. In this work, we propose a new problem, named as, \emph{affinity classification problem}. We develop a devoted machine that is embedded in a 2-dimensional cellular automaton having Moore neighborhood dependency and periodic boundary condition. Our model has affection capabilities to a converging point, all-$1$ or all-$0$ and can be characterized by four parameters $K, \phi(x), \psi(x)$ and $p$.
	Using this model, we can develop a self-healing system. We know that, because of self-healing any species can survive in the evolution. As our model has this feature and it takes decision democratically, we can say that the model is acting like a natural living system to some extent and we can conclude that the model become intelligent. 
	
	However, there are some other properties of life which an intelligent machine need to possess; we have to see if our model possess them. Similarly, here we have considered only Moore neighborhood, what kind of behavior might arise if we change the neighborhood dependency for the rules is still not seen. 
	Different other behaviors might emerge by varying the parameters of our model. And, apart from self-healing systems, our model may be useful for other several areas of application. Answers to these questions remain work of the future.
	
	\bibliographystyle{plain}
	\bibliography{References_thesis.bib}	
\end{document}